\documentstyle[aps,multicol,epsf]{revtex}

\begin{document}
\draft

\title{Influence of Island Diffusion on Submonolayer Epitaxial Growth}

\author{P. L. Krapivsky,$^{1}$
J. F. F. Mendes,$^{2}$
and S. Redner$^{1}$
}

\address{$^{1}$Center for Polymer Studies and Department of Physics, 
Boston University, Boston, MA 02215 \\
$^{2}$Centro de F\'\i sica do Porto and Departamento de F\'\i sica, 
Universidade do Porto, 4150 Porto, Portugal}

\maketitle

\begin{abstract}  
  We investigate the kinetics of submonolayer epitaxial growth which is
  driven by a fixed flux of monomers onto a substrate.  Adatoms diffuse on
  the surface, leading to irreversible aggregation of islands.  We also
  account for the effective diffusion of islands, which originates from
  hopping processes of their constituent adatoms, on the kinetics.  When the
  diffusivity of an island of mass $k$ scales as $k^{-\mu}$, the (mean-field)
  Smoluchowski rate equations predicts steady behavior for $0\leq\mu<1$, with
  the concentration $c_k$ of islands of mass $k$ varying as $k^{-(3-\mu)/2}$.
  For $\mu\geq 1$, a quasi-static approximation to the rate equations
  predicts slow continuous evolution in which the island density increases as
  $(\ln t)^{\mu/2}$.  A more refined matched asymptotic expansion reveals
  unusual multiple-scale mass dependence for the island size distribution.
  Our theory also describes basic features of epitaxial growth in a more
  faithful model of growing circular islands.  For epitaxial growth in an
  initial population of monomers and no external flux, a scaling approach
  predicts power-law island growth and a mass distribution with a behavior
  distinct from that of the non-zero flux system.  Finally, we extend our
  results to one- and two-dimensional substrates.  The physically-relevant
  latter case exhibits only logarithmic corrections compared to the
  mean-field predictions.

\end{abstract}

\begin{multicols}{2}

\narrowtext

\section{Introduction}

Submonolayer epitaxial thin film growth involves deposition of atoms onto a
substrate and diffusion of these adatoms (monomers) leading to their
aggregation into islands of ever-increasing size\cite{metois,ven}.  The
resulting island morphology and mass distribution ultimately depend on these
diffusion processes on the substrate.  While it has long been recognized that
these mass transport details are crucial to epitaxial growth \cite{metois},
its ramifications are still incompletely understood.  Part of the reason for
slow progress is that a variety of microscopic details can and do influence
the rate at which adsorbate islands move
\cite{paul,bla,bartelt,kell,voter,wang,biham,morg,wen,pai,khare,stan}.

One direction of previous investigation was based on a picture that only
monomers can diffuse while larger islands are
immobile\cite{ven,paul,bla,bartelt}.  In this case, the island density $N$
scales with time as $t^{1/3}$ before it reaches a maximum density which
scales with flux $F$ as $F^{1/3}$.  Subsequent work revealed, however, that
basic results are sensitive to minor alterations to the mass transport
mechanism.  For example, if both monomers and dimers diffuse while larger
islands remain immobile, the exponents of the time and flux dependences
change\cite{biham}.  More generally, these exponents depend on the threshold
size between mobile and immobile islands.

On the other hand, there has been increasing
appreciation\cite{morg,wen,pai,khare} that adatom hopping continues to occur
even when they are incorporated into islands of arbitrary size.  This leads
to a non-zero diffusivity of such islands.  For appropriate systems and
experimental conditions there is ample evidence that the effective island
diffusivity $D_k$ has a power-law dependence on mass
$k$\cite{morg,wen,pai,khare}.  This adatom hopping also typically leads to
islands maintaining compact shapes, an important simplifying feature for
theoretical modeling.

The goal of this paper is to determine the aggregation kinetics of compact
islands in the submonolayer regime with power law and more general
mass-dependent island diffusivities\cite{kmr}.  An essential, and at first
sight surprising feature about such compact islands is that their reactivity
depends only logarithmically on island size in two dimensions.  Accordingly,
our theoretical treatment is based on a model of point-like islands.  Such a
description should accurately describe island growth, up to these logarithmic
corrections in the low coverage limit.
 
Within the mean-field Smoluchowski rate equations, we will consider a system
with: (i) fixed monomer flux, (ii) point-like islands with diffusivity which
decays faster than inversely with island mass, and (iii) irreversible
mass-conserving coalescence of islands.  We will show that the number of
islands grows logarithmically in time and that the island mass distribution
exhibits a multiple-scale mass dependence.  Our predictions are also found to
apply to a more faithful model of epitaxial growth in which islands are
growing circles, with the radius of a mass-$k$ island proportional to
$\sqrt{k}$.  Thus the point-island model provides a useful framework to
describe submonolayer epitaxial growth and gives quantitative predictions
which are robust to variations in model parameters.

In Sec.~II, we introduce our model and discuss the applicability of the
Smoluchowski approach to epitaxial growth.  In Sec.~III, we present our main
results about the asymptotic growth of islands based on the rate equations.
Different behaviors arise depending on whether the mobility exponent $\mu$ is
smaller, larger, or equal to unity.  We also discuss the flux dependence of
the maximum island density and the time range where our theory should apply.
In Sec.~IV, we briefly discuss island growth kinetics in the absence of
external flux.  In Sec.~V, we present numerical simulations and compare
results for the epitaxial growth of point islands and growing circular
islands with radius proportional to $\sqrt{k}$.  Sec.~VI contains a summary
and discussion.  In Appendix A, we generalize our theory to one-dimensional
substrates and in Appendix B we outline a more accurate treatment for
two-dimensional substrates which accounts for the logarithmic corrections in
the reaction rate.

\section{Microscopic Model}

We consider submonolayer epitaxial growth which results from the irreversible
deposition of atoms from a gas onto a substrate and subsequent irreversible
aggregation.  Adatoms (or monomers) diffuse on the substrate and aggregate
upon colliding, creating dimers and larger islands.  We consider the class of
models for which adatoms which are already incorporated into islands can
continue to diffuse -- this often occurs along the periphery of islands (see
{\it e.g.}, Ref.~\cite{khare}).  Such adatom hopping induces an effective
diffusion in which islands of mass $k$ hop on the substrate with diffusion
coefficient $D_k$.  This microscopic mechanism also typically leads to
compact island shapes.  Whenever two islands meet and aggregate, we assume
that adatom hopping quickly causes the resulting aggregate to become compact.
We therefore treat islands as circular throughout our modeling.

Within this picture for islands, a classic way to calculate the island mass
distribution is based on the Smoluchowski rate equations\cite{drake,ernst}.
This approach requires knowledge of the rate $K_{ij}$ at which an island of
mass $i$ and an island of mass $j$ aggregate to form an island of mass $i+j$.
If the aggregation process is diffusion-controlled and islands are spherical,
this aggregation rate is given by the Smoluchowski formula $K_{ij}\sim
(D_i+D_j)(R_i+R_j)^{d-2}$ in $d$ dimensions\cite{drake}.  Here $R_i$ is the
radius of an island of mass $i$.  This formula applies for $d>2$, while in
the physically relevant case of two-dimensional substrates, there is only a
slow logarithmic dependence on island radius\cite{ernst}.  Thus a reasonable
starting point for theoretical investigation is to ignore the logarithmic
term; this significantly simplifies the resulting analysis.  In Appendix B we
will return to the two-dimensional case and show that these logarithmic
corrections do not alter our main findings but rather give rise to a
logarithmic renormalization of the monomer flux.

As discussed in the Introduction, we consider the island diffusion
coefficient to be a homogeneous function of island mass, $D_k\sim k^{-\mu}$
with $\mu$ non-negative on basic physical grounds.  In appropriate time units
the reaction rate is
\begin{equation}
\label{rate}
K_{ij}=i^{-\mu}+j^{-\mu}. 
\end{equation}
However, our approach can equally-well be applied to {\em any} functional
dependence of $D_k$ on $k$ which decays faster than $k^{-1}$ as $k\to\infty$.

For a self-contained discussion, we give a qualitative argument for the
exponent value $\mu=3/2$ for the ``periphery'' adatom hopping mechanism; this
has been obtained previously by a Langevin approach\cite{khare,gruber}.  In
periphery diffusion, adatoms on the edge of an island can hop freely to
neighboring sites on the periphery.  Consider an island of linear size $R\gg
1$.  In a time interval $\Delta t\sim R^2$ an adatom on the edge will
typically explore the entire island boundary, so this adatom will typically
be a distance $R$ from its initial location.  Hence the effective island
center-of-mass displacement is $\delta x\sim R/R^2=R^{-1}$.  If each edge
adatom diffuses independently, the total center-of-mass displacement $\Delta
x$ is the sum of $R$ independent identically distributed random variables
with variance $\delta x\sim R^{-1}$.  This implies $\Delta x\sim
\sqrt{R(\delta x)^2}\sim R^{-1/2}$.  Thus the effective center-of-mass
diffusion coefficient is $D_R\sim (\Delta x)^2/\Delta t\sim R^{-3}$, or
$D_k\sim k^{-3/2}$.  For a $d$-dimensional substrate, a straightforward
generalization of this argument gives $\mu=1+1/d$.  Similarly, in the case of
so-called ``terrace'' diffusion\cite{khare}, it is found that $D_k\sim
k^{-1}$ independent of $d$.

Before presenting detailed results, we discuss some limitations of our rate
equation approach.  As stated previously, we ignore the effect of a finite
island radius on the form of the reaction rates, as this dependence is only
logarithmic in two dimensions.  Moreover, as the coverage increases a rate
equation description based on two-body aggregation eventually breaks down.
Our approach is also inapplicable to fractal-shape islands, a situation which
can arise at low temperature\cite{fractal}.  Finally, the rate equation
description should fail below an upper critical dimension $d_c$\cite{van}
which is given by $d_c=2/(1-\mu-\lambda)$, with $\lambda$ the homogeneity
degree of the reaction kernel.  For our point-island model with kernel given
by Eq.~(\ref{rate}), $\lambda=-\mu$ and hence $d_c=2$.  Thus the relevant
two-dimensional case corresponds to the critical dimension, and logarithmic
corrections to mean-field predictions, in addition to the
previously-discussed logarithmic corrections to the reaction rate, can be
anticipated.

\section{Growth with Flux}

Consider a point-island system with flux $F$ of monomers onto the substrate.
The rate equations for the concentrations $c_k$ of islands of mass $k$ in the
presence of this steady monomer flux are
\begin{equation}
\label{smol}
{dc_k\over dt}= {1\over 2}\sum_{i+j=k}K_{ij}c_ic_j
-c_k\sum_{k=1}^\infty K_{kj}c_j+F\,\delta_{k1}.
\end{equation}
These rate equations represent a mean-field approximation in which
spatial fluctuations are neglected, and also a low-coverage
approximation, since only binary interactions are treated.  

\subsection{Steady State Regime}

Let us first consider the behavior in the steady state regime.  To analyze
this system of rate equations when the time derivative term is set to zero,
it proves convenient to use the generating function technique.  We introduce
two generating functions
\begin{equation}
\label{moments}
{\cal C}(z)=\sum_{k=1}^\infty c_k z^k, \quad
{\cal C}_\mu(z)=\sum_{k=1}^\infty k^{-\mu}c_k z^k.
\end{equation}
Multiplying Eq.~(\ref{smol}) by $z^k$, and summing over all $k$, gives
\begin{equation}
\label{mom}
{\cal C}_\mu(z){\cal C}(z)-{\cal C}_\mu(z)N
-{\cal C}(z)N_\mu + Fz=0.
\end{equation}
Here $N={\cal C}(z=1)=\sum c_k$ is the total island density
and $N_\mu={\cal C}_\mu(z=1)=\sum k^{-\mu}c_k$.

We now assume a power law asymptotic behavior for the steady state
concentration,
\begin{equation}
\label{asym}
c_k\simeq {C\over k^\tau}, 
\end{equation}
as $k\to\infty$.  For this power law to hold for all $k$, we require
$\tau>1$, so that $\sum k^{-\tau}$ converges; this leads to the
condition $\mu<1$ for the mobility exponent as shown below.  From basic
Tauberian theorems\cite{hardy}, the asymptotic form for $c_k$ in
Eq.~(\ref{asym}) induces the following power-law singularities in the
generating functions as $z\to 1$
\begin{eqnarray}
\label{gen}
{\cal C}(z) &=&N+C\Gamma(1-\tau)(1-z)^{\tau-1}+\ldots,\nonumber \\
{\cal C}_\mu(z)&=&
N_\mu+C\Gamma(1-\tau-\mu)(1-z)^{\tau+\mu-1}+\ldots,
\end{eqnarray}
where $\Gamma$ is the gamma function.  The leading constant factor in
each line is finite and coincides with the definition given in
Eq.~(\ref{moments}) if the exponent of the second term is positive.
Otherwise, the constant factor vanishes and the generating function has
a power-law divergence as $z\to 1$.  Substituting these expansions into
Eq.~(\ref{mom}) and matching the leading behavior in $(1-z)$ leads to
the decay exponent $\tau=(3-\mu)/2$.  The condition for a steady state
to occur, $\tau>1$, thus imposes an upper bound on the mobility
exponent, $\mu<1$.  From matching the leading behavior in $(1-z)$, the
constant $C$ may also be determined, from which the island mass
distribution in the steady-state regime $0\leq\mu<1$ is
\begin{equation}
\label{steady}
c_k\simeq \sqrt{F(1-\mu^2)\cos(\pi\mu/2)\over 4\pi}
\,k^{-(3-\mu)/2}.
\end{equation}

In the special case of constant reaction rate, $\mu=0$, one can find
a complete time-dependent solution.  In this case, the rate equations are
\begin{equation}
\label{solv}
{dc_k\over dt}=
\sum_{i+j=k}c_ic_j-2c_kN+F\delta_{k1}
\end{equation}
By summing these equations over $k$, one obtains the associated equations for
the density,
\begin{eqnarray}
\label{N}
{dN\over dt}&=&-N^2+F,
\end{eqnarray}
and for the generating function,
\begin{eqnarray}
\label{cal}
{d{\cal C}(z,t)\over dt}={\cal C}(z,t)-2{\cal C}(z,t)N(t)+Fz.
\end{eqnarray}
Solving these equations for initially clean surface, $c_k(0)=0$,
we obtain the time-dependent island density
\begin{equation}
\label{solN}
N(t)=\sqrt{F}\,\tanh\left(t\sqrt{F}\right), 
\end{equation}
and the generating function
\begin{equation}
\label{solC}
{\cal C}(z,t)=N(t)-\sqrt{F(1-z)}\tanh\left(t\,\sqrt{F(1-z)}\right).
\end{equation}

Expanding the generating function ${\cal C}(z,t)$ in a series in $z$, gives
the concentrations $c_k(t)$.  Relatively simple results are obtained in the
long time limit
\begin{equation}
\label{solck}
c_k(t)=\sqrt{F\over 4\pi}\,{\Gamma\left(k-{1\over 2}\right)
\over \Gamma(k+1)}.
\end{equation}
This exact solution of Eq.~(\ref{solck}) agrees with the asymptotic solution
of Eq.~(\ref{steady}).  Note that islands of sizes comparable with
$(1-z)^{-1}$ or smaller, give dominant contribution to the generating
function, ${\cal C}_0(z,t)=\sum_{k\geq 1}z^kc_k(t)$, while bigger islands
provide an asymptotically negligible correction.  Eq.~(\ref{solC}) also shows
that the crossover from a time-dependent to saturated behavior takes place
when $t\sqrt{1-z}\sim 1$.  Together with $k\sim (1-z)^{-1}$ this implies that
the steady state is established for small islands whose size is in the range,
$k\ll t^2$.  For general mobility exponent, one physically expects a similar
crossover to steady behavior when $k\sim t^\zeta$, with mass cutoff exponent
$\zeta$ dependent on $\mu$.  To determine $\zeta$, we use the physical
condition that the total mass on the substrate $\sum kc_k(t)$ equal $Ft$
together with the steady state asymptotics of Eq.~(\ref{steady}).  This gives
\begin{equation}
\label{zeta}
\sum_{k=1}^\infty kc_k(t)\sim 
\sum_{k=1}^{t^\zeta}k^{(\mu-1)/2}\sim t^{(\mu+1)\zeta/2}\sim t,
\end{equation}
that is, $\zeta=2/(\mu+1)$.

\subsection{Continuous Island Evolution}

For sufficiently large mobility exponent, $\mu\geq 1$, continuous evolution
can be anticipated.  Indeed, in the previous subsection we showed that
assuming a steady state leads to $\mu<1$.  Additionally, continuous evolution
is known to occur when $\mu=\infty$, that is, in the extreme case of mobile
monomers and {\it immobile} islands.  This case has been treated analytically
\cite{ven,paul,bla,bartelt} and power-law growth in the number of islands,
$N(t)\simeq (3F^2 t)^{1/3}$, was found.  Moreover, in the scaling region
\begin{equation}
\label{region}
k\to \infty, \quad t\to \infty, \qquad {\rm with}\quad 
x={2k\over \left(Ft^2\right)^{1/3}}<1,
\end{equation}
the island mass distribution approaches the scaling form
\begin{equation}
\label{scaling}
c_k(t)\simeq \left({F\over 3t}\right)^{1/3}(1-x)^{-1/2}, 
\qquad {\rm when} \quad x<1,
\end{equation}
while for $x>1$ the island concentrations are negligible. 

We therefore expect that for all $1\leq \mu\leq\infty$, the island mass
distribution does not reach a steady state.  The marginal case of $\mu=1$ is
interesting since it corresponds to experimentally relevant case of terrace
diffusion\cite{khare}.  As discussed in the previous section, such values of
the mobility exponent $\mu$ naturally appear for different microscopic
mass-transport mechanisms.

Our primary results are that when $\mu$ is strictly greater than unity but
still finite, 
\begin{equation}
\label{log}
N(t)\simeq \sqrt{F}\,
\left[{\sin(\pi/\mu)\over \pi}\,\,\ln \left(t\sqrt{F}\right)\right]^{\mu/2}, 
\end{equation}
while the concentration of islands of mass $k$ decays in
time as
\begin{equation}
\label{clog}
c_k(t)\sim \sqrt{F}\,(k!)^\mu\,
\left[\ln \left(t\sqrt{F}\right)\right]^{-\mu(2k-1)/2}
\end{equation}
for sufficiently small islands, $k\ll \ln(t\sqrt{F})$.  It is remarkable that
these logarithmic dependences, a feature which generally signals marginal
behavior, occurs in the entire regime $1<\mu<\infty$.

Our approach that leads to Eqs.~(\ref{log}) and (\ref{clog}) is based on the
physical picture that the system is slowly evolving because the growth of
islands by aggregation is substantially counterbalanced by the input of
monomers.  These competing effects lead to nearly time-independent island
concentrations over an ``inner'' size range, while more
strongly-time-dependent behavior occurs in an ``outer'' range which extends
to the largest islands.  In the inner region, the picture of near balance
between aggregation and input motivates the use of the quasi-static
approximation, where the time derivative in Eq.~(\ref{smol}) is initially
neglected, from which the island concentrations as a function of the total
concentration of islands can be obtained.  The dependence of the island
concentrations on time is then determined by the condition that the total
mass in the system is proportional to $t$.  The validity of this approach may
be verified {\it a posteriori}, where the logarithmic dependences in
Eqs.~(\ref{log}) and (\ref{clog}) imply that the temporal derivatives in the
Smoluchowski rate equations are asymptotically negligible.

Within this quasi-static framework, Eqs.~(\ref{smol}) become
\begin{eqnarray}
\label{sm}
0&=&1-c_1\left(N+N_\mu\right), \nonumber \\
0&=&\frac{1}{2}\sum_{i+j=k}\left({i^{-\mu}}+{j^{-\mu}}\right)c_ic_j
   -c_k\left(k^{-\mu}N+N_\mu\right).
\end{eqnarray}
By summing Eqs.~(\ref{sm}) over all $k$, the total island density in the
quasi-static limit obeys
\begin{equation}
\label{nqs}
0=1-NN_\mu
\end{equation}
Eqs.~(\ref{sm}) and (\ref{nqs}) have been non-dimensionalized by the scale
transformation $c_k\to \sqrt{F}\,c_k$.  We also scale the time variable by
$t\to t/\sqrt{F}$, so that the mass density obeys
\begin{equation}
\label{mass}
\theta(t)=\sum_{k=1}^\infty kc_k(t)=t.
\end{equation}

Eq.~(\ref{nqs}) immediately gives $N_\mu=N^{-1}$, and then from the first of
Eqs.~(\ref{sm}), $c_1\simeq 1/N$.  The remainder of Eqs.~(\ref{sm}) may then
be solved recursively.  By writing the first few of these equations, it is
evident that the dominant contribution to $c_k$ is the term in the quadratic
product which is proportional to $c_1c_{k-1}$.  Keeping only this
contribution, the resulting recursion may be solved straightforwardly to
yield
\begin{eqnarray}
\label{soln}
c_k&\simeq&{1\over N}\,\left[\prod_{j=2}^k
{j^\mu/N^2\over{1+j^\mu/N^2}}\right]\,
\prod_{j=1}^{k-1}\left(1+j^{-\mu}\right) \nonumber\\
&\equiv&{1\over N}\prod_{j=2}^k B_j\prod_{j=1}^{k-1} b_j.
\end{eqnarray}
Since the factors $B_j\ll 1$ for $j^\mu\ll N^2$, while $B_j\to 1$ for
$j^\mu\gg N^2$, this implies that $c_k$ is a rapidly decreasing function
of $k$ for $k\ll N^{2/\mu}\equiv\kappa$ and then converges to a finite
value $\rho$ for $k>\kappa$.

To compute $c_k$, first note that for $\mu>1$ the product $\prod_j b_j$
converges, so that it may treated as constant.  We then write the second
product as the exponential of a sum and take the continuum limit.  This
leads to
\begin{eqnarray}
\label{ck-cont}
c_k&\sim&{1\over N}\exp\left
[\sum_{j=2}^k\ln\left({j^\mu/N^2\over{1+j^\mu/N^2}}\right)\right]\nonumber\\
 &\simeq& {1\over N}\exp\left[-N^{2/\mu}\int_0^x\ln(1+w^{-\mu})\,dw\right],
\end{eqnarray}
where $w=j/N^{2/\mu}$ and $x=k/N^{2/\mu}$.  This form has two slightly
different asymptotic behaviors depending on whether $\mu$ is strictly greater
than or equal to 1.  For $\mu>1$, the monotonically increasing integral in
Eq.~(\ref{ck-cont}) converges as $x\to\infty$.  Thus $c_k$ decreases as a
function of $k$ until a threshold value $\kappa\simeq N^{2/\mu}$, beyond
which $c_k$ remains constant with a value $\rho$ determined by taking the
upper limit of the integral as infinite.  Hence
\begin{equation}
\label{cstar}
c_k\to \rho\sim {1\over N}\exp\left[-A_\mu\,N^{2/\mu}\right], 
\end{equation}
with $A_\mu=\int_0^\infty\ln(1+w^{-\mu})\,dw=\pi/\sin(\pi/\mu)$.  Within the
quasi-static approximation, this constancy in $c_k$ should persist over the
range, $\kappa\alt k\ll K$.  Here $K$ defines an upper limit for the
``inner'' regime where the quasi-static approximation remains valid.
Physically, this limit corresponds to the size range where islands are only
beginning to form.

Because of the importance of the temporal decay of the densities in the outer
size range, $k\gg\kappa$, the quasi-static approximation is inadequate and an
alternative is needed.  We therefore separately account for the distribution
of these ``raw'' (evolving) islands and then perform a matched asymptotic
expansion\cite{nayfeh}, to join the inner ($k\ll K$) quasi-static solution of
``ripe'' islands to the outer ($k\gg\kappa$) solution of raw islands in the
overlap region $\kappa\ll k\ll K$.  In the outer region, the asymptotically
dominant terms in the rate equations are\cite{inf}
\begin{equation}
\label{sm-raw}
{dc_k\over dt}=c_1\left(c_{k-1}-c_k\right).
\end{equation}
Since raw islands are large, we employ the continuum limit of
Eq.~(\ref{sm-raw}),
\begin{equation}
\label{raw}
\left({\partial \over \partial T}+{\partial \over \partial k}
-{1\over 2}\,{\partial^2 \over \partial k^2}
\right)\,
c_k(T)=0,
\end{equation}
where $T=\int_0^t dt'\,c_1(t')$.  If we neglect the diffusive term, the
solution of the resulting wave equation is $c_k(T)=f(T-k)$, with $f$ an
arbitrary function.  Matching with the inner solution determines $f$ and gives
\begin{equation}
\label{ck-raw}
c_k(T)=\rho(T-k),
\end{equation}
which thus provides the time dependence of the island concentrations once $T$
has been determined.  Also, Eq.~(\ref{ck-raw}) provides the estimate
$K\approx T$ for the cutoff size beyond which essentially no islands exist.

To close the solution, we now determine the time dependence of the island
density $N(t)$.  To accomplish this, we use the sum rules for the mass and
island densities, $\sum kc_k=M\equiv t$ and $\sum c_k=N$.  In the continuum
limit these become
\begin{equation}
\label{sum}
t=\int_0^T du\,(T-u)\rho(u), \quad
N=\int_0^T du\,\rho(u).
\end{equation}
The dominant contribution to these integrals arises from large (raw) islands
in the outer region.  The second relation also gives ${dN\over dT}=\rho(T)$.
Combining this with Eq.~(\ref{cstar}) yields
\begin{equation}
\label{rho}
\rho(T)\sim {(\ln T)^{\mu/2-1}\over T}.
\end{equation}
The raw island mass distribution is then given by Eqs.~(\ref{ck-raw})
and (\ref{rho}) as,
\begin{equation}
\label{cT-raw}
c_k(T)\sim {\left[\ln (T-k)\right]^{\mu/2-1}\over T-k}.
\end{equation}
Notice that the singularity when $k=K=T$.  However in the mass range $T-k\sim
\sqrt{T}$, this singularity is smoothed out by the diffusive term in
Eq.~(\ref{raw}).  Thus instead of the singularity at $k=K$, the density of
raw islands reaches a peak value of the order of $t^{-1/2}$ and then rapidly
decreases for larger $k$.

Using Eq.~(\ref{rho}), we can now estimate the left integral in
Eq.~(\ref{sum}) and find that the ratio of the first to the second term
scales as $\ln T$.  Keeping only the dominant contribution and using the
right integral of (\ref{sum}) gives $T\simeq t/N$.  Thus the cutoff size is
given by
\begin{equation}
\label{K}
K(t)\simeq {t\over N}\sim {t\over{(\ln t)^{\mu/2}}}.
\end{equation}
Parenthetically, the basic relation between the real and modified times,
$T=\int_0^t dt'c_1(t')$, together with $c_1\cong 1/N$ also leads to $T\simeq
t/N$ and thus to Eq.~(\ref{K}).

In real time the island density becomes
\begin{equation}
\label{lognew}
N(t)\simeq 
\left\{A_\mu^{-1}\,
\ln\left[t\left({\ln t\over A_\mu}\right)^{1-3\mu/2}\right]\right\}^{\mu/2},
\end{equation}
where $A_\mu=\pi/\sin(\pi/\mu)$.  Upon neglecting the logarithmic temporal
factor inside the logarithm, our basic result quoted in Eq.~(\ref{log}) is
recovered.

\subsection{The Case $\mu=1$}

In the borderline case $\mu=1$, subtler nested logarithmic behavior arises,
as reflected by a singularity in Eq.~(\ref{log}) that appears upon formal
continuation to $\mu\to 1$.  The analysis closely parallels the case $\mu>1$
and we just outline the main results.

In the inner region, the general approach of the previous subsection applies
up to Eq.~(\ref{soln}).  However, when $\mu=1$, the product
$\prod_{j=1}^{k-1} b_j=\prod_{j=1}^{k-1} (1+j^{-1})$ in Eq.~(\ref{soln}) now
equals $k$, {\it i.e.}, it diverges.  Second, the term $c_2c_{k-2}$ in
addition to $c_1c_{k-1}$ contributes to the asymptotic behavior.  Due to this
latter attribute, the recursion relation for $c_k$ becomes
\begin{equation}
\label{ck-mar}
{c_k\over k}\,{1+k/N^2\over k/N^2 }={c_{k-1}\over k-1}
+{c_{k-2}\over k-2}\,{1\over N^2}.
\end{equation}
We seek a solution for $c_k$ in the form of Eq.~(\ref{soln}).  Thus we
write
\begin{equation}
\label{ck-marg}
c_k\sim {\cal C}_k\,{k\over N}\left[\prod_{j=2}^k
{j/N^2\over{1+j/N^2}}\right],
\end{equation}
where the factor ${\cal C}_k$ accounts for the additional term in
Eq.~(\ref{ck-mar}).  Substituting into Eq.~(\ref{ck-mar}) gives the recursion
formula for this correction factor ${\cal C}_k$
\begin{equation}
\label{Bk}
{\cal C}_k={\cal C}_{k-1}
+{\cal C}_{k-2}\left({1\over k-1}+{1\over N^2}\right).
\end{equation}
These coefficients are slowly varying in $k$ when $k\gg 1$ and we may
treat $k$ as continuous in this asymptotic regime.  Eq.~(\ref{Bk}) then
becomes a differential equation whose solution is ${\cal C}_k\sim
k\,e^x$ (with $x=k/N^2$).  Consequently,
\begin{equation}
\label{ck-margin}
c_k\sim {k^2\over N}\,\exp\left[x
-N^2\int_0^x \ln\left(1+w^{-1}\right)\,dw\right].
\end{equation}
Thus for $\mu=1$, the concentration $c_k$ decreases rapidly in $k$ for
$k\ll N^2$ and reaches a minimum at $k=N^4$ whose value is
\begin{equation}
\label{ck-star-marg}
\rho\sim \exp\left[-N^2\ln N^2\right].
\end{equation}
In the outer region, the general results of the previous subsection,
Eqs.~(\ref{sm-raw}) through (\ref{sum}), are still valid.  Thus combining 
(\ref{ck-star-marg}) with ${dN\over dT}=\rho(T)$ we solve for 
the total island
concentration to obtain
\begin{equation}
\label{2log}
N(t)\sim \sqrt{\ln t\over \ln(\ln t)}.
\end{equation}
Eq.~(\ref{ck-marg}) together with ${\cal C}_k\sim k$ implies that the
concentration of islands for mass $k\ll N^2$ is
\begin{equation}
\label{3log}
c_k(t)\sim {(k+1)!\over N^{2k-1}}
\sim (k+1)!\,\left[{\ln(\ln t)\over \ln t}\right]^{k-1/2}.
\end{equation}
Then the island mass density is approximately constant, $c_k(T)\simeq
\rho(T)$, where
\begin{equation}
\label{4log}
\rho(T)\sim T^{-1}(\ln T)^{-1} [\ln(\ln T)]^{-1}.  
\end{equation}
This holds over the range $\kappa\alt k\ll K$ with $K(t)\sim
t\,\sqrt{\ln(\ln t)/\ln t}$.  Finally, the raw island density given by
Eqs.~(\ref{ck-raw}) and (\ref{4log}) holds up to a mass cutoff $K(t)$.

\subsection{Maximum Island Density}

To apply our results to real submonolayer epitaxial systems, note that in the
submonolayer regime, the coverage must be small, that is, $M\equiv Ft \ll 1$.
On the other hand, the asymptotic predictions of our theory apply when
$t\sqrt{F}\gg 1$.  Consequently, our results should be valid in the time
range $F^{-1/2}\ll t\ll F^{-1}$.  Since the (dimensionless) flux $F$ is
typically small in epitaxy experiments, the time range over which our theory
should apply is correspondingly large.  A commonly employed connection
between theory and experimental results is to determine the maximum island
density at the end of the submonolayer regime, $t_{\rm max}\sim F^{-1}$.  The
conventionally-quoted result is that this maximum density scales as a power
of the flux\cite{ven,old}
\begin{equation}
\label{Nmaxold}
N_{\rm max}\sim F^{\chi} 
\end{equation}
with the flux exponent typically in the range ${1\over 3}\leq \chi\leq
{1\over 2}$\cite{biham,stan,mo,ern,vill,wolf}.  While most theoretical
studies predict $\chi<1/2$\cite{biham,stan,vill,wolf}, ({\it e.g.,} for point
immobile islands $\chi=1/3$), the value $\chi=1/2$ has been observed
experimentally, see Ernst {\it et al.}\cite{ern}.  Our analysis predicts that
the maximum island density attains the value
\begin{equation}
\label{Nmax}
N_{\rm max}\sim F^{1/2}[\ln(1/F)]^{\mu/2},
\end{equation}
and that $\chi=1/2$ is generic and applies to any model where island
diffusion leads to continuous evolution.  Previous theoretical work has
focused on somewhat pathological models where only a few (the smallest)
island species could diffuse.  In such models, the exponent $\chi$ is
sensitive to the cutoff size between mobile and immobile islands.
Fortunately, the generic situation is simpler and the asymptotic flux
dependence of $N_{\rm max}$ is universal.  However, there do exist
logarithmic factors which are non-universal, as they depend on the mobility
exponent.  For example, the logarithmic factor in Eq.~(\ref{Nmax}) gives an
effective exponent, when plotting $N_{\rm max}$ versus $F$, $\chi_{\rm
  eff}={1\over 2}\left[1-\mu\, {\ln(\ln(1/F))\over \ln(1/F)}\right]$, which
is {\em smaller} than 1/2.  Indeed, for a fixed flux, the effective flux
exponent is a decreasing function of $\mu$.  Such a feature should be taken
into account in the interpretation of experimental and numerical data.

\section{Growth without Flux}

We now consider epitaxial growth for point islands with a non-zero initial
monomer density, reaction rate $K_{ij}=i^{-\mu}+j^{-\mu}$, and no subsequent
monomer flux.  Such a system has been extensively investigated within the
framework of irreversible aggregation\cite{meakin,oshanin,sholl,sire,kandel},
as well as in theoretical studies of point-island aggregation with immobile
islands\cite{paul,sander}.  More realistic examples of the latter system have
also been investigated numerically and experimentally in recent studies (see
\cite{pai,sholl2} and references therein).

We give here a simple argument for the asymptotic form of the island size
distribution.  This argument is based on first solving for the island and
monomer densities, $N(t)$ and $c_1(t)$, respectively, and then using
scaling\cite{ernst} to infer the asymptotics of the distribution.  According
to the scaling ansatz, the asymptotic island-mass distribution should have
the following form
\begin{equation}
\label{sc}
c_k(t)\simeq N^2G(x), \quad x=kN,
\end{equation}
for finite $x$.  The constraints $\int dx\, G(x)=1$ and $\int dx\,x\,
G(x)=\theta$ automatically enforce the conditions $\sum c_k(t)=N(t)$ and mass
conservation $\sum kc_k(t)=\theta$.  Starting with the exact rate equation
for $N(t)$,
\begin{equation}
\label{rateN}
{dN\over dt}=-{1\over 2}\,\sum_{i=1}^\infty \sum_{j=1}^\infty 
K_{ij}c_ic_j,
\end{equation}
we substitute the scaling ansatz Eq.~(\ref{sc}) to obtain $\dot N\sim -N^{2+\mu}$
whose solution is
\begin{equation}
\label{Nt}
N(t)\sim t^{-1/(1+\mu)}. 
\end{equation}

Consider now the rate equation for the monomer density, $\dot
c_1=-Kc_1(N+N_\mu)$.  As shown by Eq.~(\ref{nqs}), we may neglect the second term
and integrate the resulting equation to yield
\begin{equation}
\label{c1}
c_1(t)\sim \exp\left[-t^{\mu/(1+\mu)}\right].
\end{equation}
Together with Eqs.~(\ref{sc}) and (\ref{Nt}), (\ref{c1}) predicts that the
asymptotic behavior for the scaling function in the small-$x$ limit is
$\exp(-1/x^\mu)$ .  On the other hand, for the entire class of reaction
kernels of the form $K_{ij}=i^{-\mu}+j^{-\mu}$, the scaling function $G(x)$
decays exponentially\cite{ernst} in $x$ for large $x$.  Combining these
results gives the asymptotic forms of the island size distribution in the
absence of monomer flux
\begin{equation}
\label{mf}
G(x)\sim \cases{e^{-1/x^\mu}, &   $x\downarrow 0$,\cr
                e^{-x},   &       $x\uparrow \infty$. \cr}
\end{equation}

To determine the validity of these mean-field predictions, let us consider
the kinetics of this process in general spatial dimension.  To this end, we
determine the reactivity of an arbitrary cluster by a dimensional argument.
Consider an island of radius $R$ which diffuses with the diffusion
coefficient $D$ in $d$-dimensional space.  During a time interval $t$ this
particle traces out the so-called Wiener sausage whose volume is\cite{feller}
\begin{equation}
\label{wiener}
V_d(t)\sim \cases{Dt\,\left[\ln(Dt/R^2)\right]^{-1}, & $d=2$,\cr
                  Dt R^{d-2},                        & $d>2$. \cr}
\end{equation}
For epitaxial growth with no flux, $R$ corresponds to a (growing) average
island radius, and $D$ to the diffusivity of a mass-$k$ island, $D\sim
k^{-\mu}\sim R^{-d\mu}$.  Clearly, all monomers initially within a Wiener
sausage will aggregate into a typical island by time $t$.  Consequently, the
initial coverage of the substrate $\theta$ is given by $R^d\sim \theta
V_d(t)$.  This, together with Eq.~(\ref{wiener}), gives $R^2\sim Dt$ in {\it
  all} dimensions $d\geq 2$.  Finally, combining $R^2\sim Dt$ and $D\sim
R^{-d\mu}$ leads to
\begin{equation}
\label{Nd}
R\sim t^{1/(2+d\mu)},\quad
N\sim t^{-d/(2+d\mu)}.
\end{equation} 

In two dimensions, the rate equation predictions (Eq.~(\ref{Nt})) and the
above heuristic argument agree.  Thus in the absence of a monomer flux, and
under the assumption of compact islands, the two sources of logarithmic
corrections -- a logarithmic dependence of the reaction rate on island radius
and the fact that the system is at the critical dimension -- evidently cancel
each other.  Therefore the Smoluchowski rate equations appear to be
asymptotically correct in two dimensions.

\section{Numerical Results}

We have performed detailed Monte Carlo simulations for island growth for two
models of epitaxial growth.  We first consider point islands, in which
single-site islands hop with equal probability to any lattice site (diffusion
on a complete graph) and aggregate whenever two islands occupy the same site.
This corresponds to the mean-field limit and thus provides a direct test of
some of the delicate approximations made in our theoretical analysis.  The
second model is a more faithful description in which circular island
``droplets'' with radius proportional to the square-root of the island mass
diffuse to nearest-neighbor sites on a two-dimensional lattice.  Whenever
there is overlap of two islands, they immediately coalesce to a single island
which is centered at the initial position of the larger island.  After each
coalescence, a test is made to determine if additional overlaps have been
created.  All such higher-order coalescences are performed until all overlaps
are resolved.  We treat here only the case with external flux. since growth
without flux has already been investigated numerically (see {\it e.g.},
\cite{kandel} and references therein).

The point-island model is relatively easy to implement.  Additionally, this
model has a technical advantage over growing droplets in that the
submonolayer description applies over a longer time range.  For 
point islands, the submonolayer regime is defined by the criterion $N\ll
1$, which is considerably less stringent than the criterion $Ft\ll 1$
appropriate for the droplet-island model.  Therefore our theoretical results
may be compared with the point-island simulations in the time range
$F^{-1/2}\ll t\ll \exp\left(-F^{-1/\mu}\right)$ and with droplet-island
simulations in the time range $F^{-1/2}\ll t\ll F^{-1}$.

\subsection{Point Islands}

Due to the point-like nature of islands and the equiprobable hopping to any
site of the system, the simulations correspond directly to the Smoluchowski
rate equations.  Simulations of point islands were performed on a graph
of $L$ sites with the following time evolution.  At any 
stage, the number of deposition events per unit time is $FL^2$, while the
number of aggregation events per unit time is $L_{\rm occ}^2$.  Here $L_{\rm
  occ}$ is the number of occupied sites in the system.  In a microscopic
event, deposition is chosen with probability $r_{\rm dep} = FL^{2}/(FL^{2}
+L_{\rm occ}^{2})$ and aggregation is chosen with probability $r_{\rm
  agg}=1-r_{\rm dep}$.  In a deposition event, a monomer is deposited onto a
randomly chosen vacant site.  We checked that restricting the deposition onto
unoccupied sites only does not alter our results.  In an aggregation event, two
sites were chosen randomly from the list of occupied sites.  If these sites
contain islands of mass $i$ and $j$, aggregation occurs with probability
$\left(i^{-\mu}+j^{-\mu}\right)/2$.  Time is then increased by $\Delta t =
1/(FL^{2}+L_{\rm occ}^{2})$, and the procedure is repeated.

To test our algorithm, we considered first the extreme case of immobile
islands, $\mu=0$, where a complete time-dependent analytical solution is
available (\ref{solC}).  Simulations were performed for an initially empty
system with $L$ between $10^3$ and $10^4$.  We found excellent agreement
between numerical and theoretical predictions, both for the steady state and
transient characteristics.  More generally for $\mu<1$, our simulations
showed that the system reaches a steady state with the observed steady state
characteristics in good agreement with the theoretical prediction of
Eq.~(\ref{steady}).

\begin{figure}
\epsfxsize=60mm
\epsfysize=60mm
\centerline{\epsffile{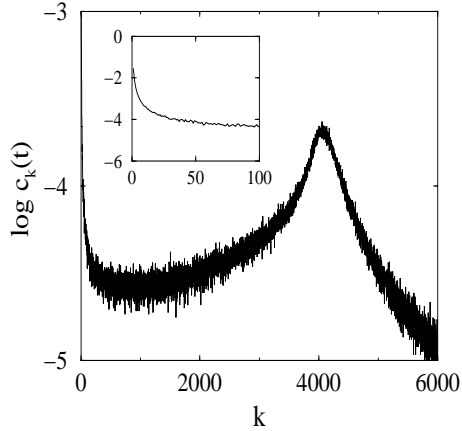}}
\caption{Semilogarithmic plot of $\ln c_k(t)$ vs $k$ at $t\approx 22000$ for
  $\mu=1.5$ based on simulations of point islands.  The data is based on 5000
  realizations of an initially empty system of 2000 sites with $F=0.05$.
  Notice the existence of an inner scale $k\protect\alt\kappa\approx 50$
  (inset) and an outer scale for $k\approx K\approx 4000$. }
\end{figure}

\begin{figure}
\epsfxsize=60mm
\epsfysize=60mm
\centerline{\epsffile{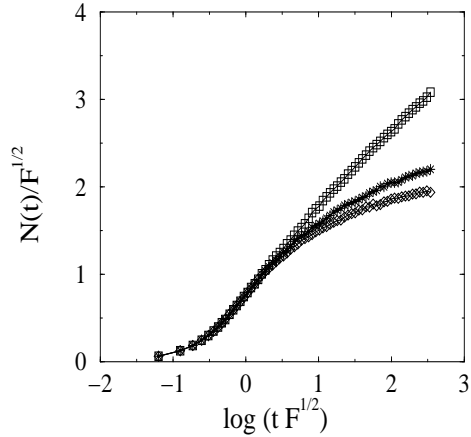}}
\caption{Plot of $N(t)/\sqrt{F}$ {\it vs}. $\log\left(t\sqrt{F}\right)$ for 
  $\mu=1.2$ ($\Diamond$), $\mu = 1.4$ ($\star$), and $\mu = 2.0$ ($\Box$).
  The simulation parameters are the same as in Fig.~1.}
\end{figure}

For $\mu>1$, the total number of islands is found to increase indefinitely
and the island mass distribution substantively agrees with our theoretical
predictions (see Fig.~1).  Indeed, the island mass distribution $c_k(t)$
sharply decreases for small mass (ripe islands), and then increases in a mass
range which corresponds to raw islands.  In the proximity of $k\approx K$,
there is a peak in $c_k$ as predicted by our description based on matched
asymptotic expansions.  Our data for the time dependence of the total island
density (Fig.~2) are consistent with $N(t)$ growing as a power of $\ln t$, as
predicted by Eq.~(\ref{log}), but with somewhat smaller exponent than
$\mu/2$.  We consider our data sufficient to exclude a power-law time
dependence of the island density.  However, impractically long simulation
would be needed to determine the exponent of the logarithmic factor in
Eq.~(\ref{log}).

\subsection{Circular Island Droplets}

We now consider simulations of compact growing circular islands which we term
as droplets.  We additionally assume that islands are always centered on
sites of the square lattice.  In the simulations, we consider a system of
size $L^2$ to which we add monomer droplets of radius $r_0= 0.495$ to
guarantee that adjacent monomers do not overlap.  Monomers, however, can
overlap with bigger droplets.  When two droplets of radii $r_1$ and $r_2$
overlap they coalesce to form a droplet of radius
$\sqrt{r_{1}^{2}+r_{2}^{2}}$ which is located at the center of the larger of
the two initial coalescing droplets.  As the coverage increases, multiple
coalescences become increasingly probable and we therefore check for new
island overlaps after each coalescence event and perform additional
coalescences and continue to resolve all additional overlaps, if needed.

In the time evolution, a microscopic process, either deposition onto any
system site or diffusion to a nearest-neighbor site is chosen with respective
probabilities $p_{\rm dep} = F/(F + N)$ and $p_{\rm dif} = 1-p_{\rm dep}$.
We then test for and perform all possible subsequent coalescences after each
event.  After completion of this microscopic event the time is incremented by
$\Delta t = (FL^{2} +NL^{2})^{-1}$ and the process is repeated.

\begin{figure}
\epsfxsize=60mm
\epsfysize=60mm
\centerline{\epsffile{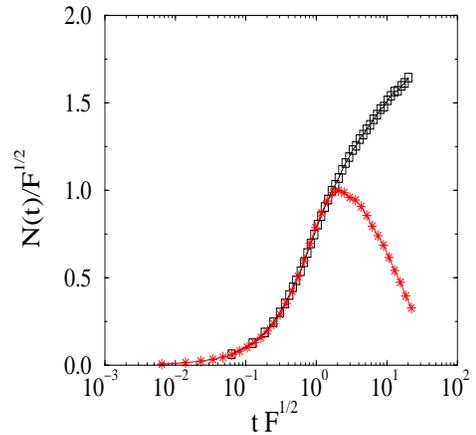}}
\caption{Plot of $N(t)/\sqrt{F}$ versus $t\sqrt{F}$ for $\mu=1.2$ and 
  $F=0.001$ for point ($\Box$) and growing circular islands
  ($\star$). }
\end{figure}

Fig.~3 compares the time evolution of point and growing circular islands.
For $t<1/\sqrt{F}$, the plots of $N(t)$ versus $t$ for the two processes
coincide, thus indicating that the point-island model provides an excellent
early-time description for the more realistic model of growing droplets.
However, when $t>1/\sqrt{F}$, multi-body aggregation starts to become
important and the density of droplets decreases with time while the density
of point islands continues to grow.  Thus the point-island model is a
suitable starting point to interpret simulational and experimental data of
epitaxial growth\cite{zan}.

A visualization of the aggregation of circular droplets at a relatively late
stage is shown in Fig.~4.  Here, many-body effects become significant and a
coalescence of two large droplets can lead to a large avalanche of coalescence
events.

\begin{figure}
\epsfxsize=60mm
\epsfysize=60mm
\centerline{\epsffile{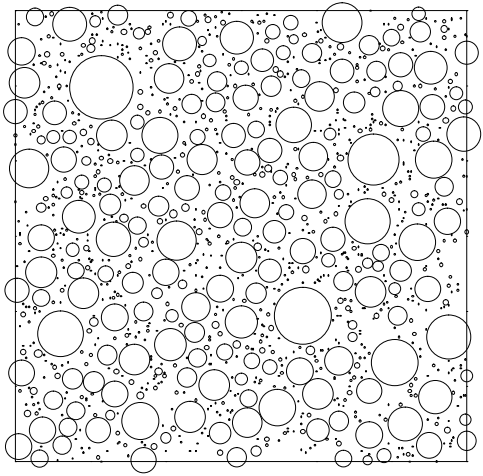}}
\smallskip
\caption{Central portion of a typical droplet configuration for $\mu =1.2$ 
  and $F=0.001$ for a system of size $500\times 500$.}
\end{figure}

In Fig.~5 we plot the dependence of the maximum island density as a function
of the flux.  Using the power-law form of Eq.~(\ref{Nmaxold}) leads to a
best-fit value $\chi\approx 0.43$, while taking into account the logarithmic
correction given by Eq.~(\ref{Nmax}) gives $\chi\approx 0.50$, in excellent
agreement with our theory.  However, using the additional flux
renormalization from our more accurate treatment of the reaction rate
(Appendix B) to fit the data leads to the somewhat different exponent
estimate of $\chi\approx 0.53$.

\begin{figure}
\epsfxsize=60mm
\epsfysize=60mm
\centerline{\epsffile{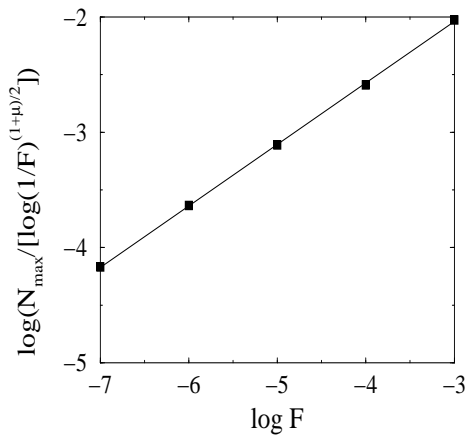}}
\caption{Plot of $\log(N_{\rm max}/[\log(1/F)]^{(1+\mu)/2})$ 
versus $\log(F)$ for $\mu =1.2$. The slope of the line is $0.53$.  }
\end{figure}

We have also performed simulations of two-dimensional fractal islands.
Specifically, when two monomers occupy neighboring sites they stick
irreversibly to form a dimer.  This process continues and leads to the
formation of fractal islands which hop as a rigid unit.  Simulation of this
process is simpler than in the growing droplet model because coalescences
subsequent to the primary event cannot occur.  This model has been studied
earlier\cite{stan} and our respective results agree.  For example, for the
case $\mu=1.2$ we obtain $\chi\approx 0.43$ if we fit the data for $N_{\rm
max}$ versus $F$ to power-law behavior.  Our interpretation for this exponent
value differs, however, from that of Ref.~\cite{stan}.  Indeed, they suggest
that there might be a deep connection to a point-island model where both
monomers and dimers diffuse and larger islands are immobile, since the
exponent in that model, $\chi=2/5$\cite{vill}, is close to their simulation
results.  We believe that there is no connection between these two models and
the exponent observed by simulation is just an effective value whose
asymptotic value is governed by the logarithmic correction in
Eq.~(\ref{Nmaxnew}).  Our theory predicts universal flux exponent $\chi=1/2$,
and the difference with observed {\em effective} flux exponent is due to the
logarithmic correction.  When this feature is taken into account, a value for
$\chi$ very close to our prediction is found.
 
\section{Summary and Discussion}

We have investigated the kinetics of submonolayer epitaxial growth within a
simple model which incorporates basic physical features of epitaxial growth
-- deposition, island diffusion, and aggregation.  We have shown that our
model displays universal kinetics -- up to logarithms -- as the governing
exponents associated with the time and flux dependence of observables are
detail independent.  This is in contrast to the behavior exhibited by models
where islands below a cutoff size are mobile while larger islands are
immobile.  In this case, characteristic exponents depend on this mass
threshold.

We analyzed in detail the situation where the effective diffusivity $D_k$ for
islands of mass $k$ is $D_k\propto k^{-\mu}$.  Such a diffusivity arises,
{\it e.g.}, in periphery diffusion, where an adatom on the edge of an island
detaches, hops to a neighboring site on the edge, and then re-attaches to the
island.  This mechanism also causes islands to be compact.  A Smoluchowski
approach shows that the reaction rate between two islands of mass $i$ and $j$
is $K_{ij}\propto i^{-\mu}+j^{-\mu}$ multiplied by a factor which depends
logarithmically on their radii in two dimensions.  This weak dependence
implies that a model with point-like diffusing islands should be
quantitatively accurate when applied to the submonolayer regime.  The net
effect of the logarithmic factor is the flux renormalization, $F\to
F\ln(1/F)$, as demonstrated in Appendix B.

For mobility exponent $0\leq \mu<1$, there is a steady state, with the
concentration of islands of mass $k$, $c_k\propto k^{-(3-\mu)/2}$.  For
$\mu\geq 1$, logarithmic time dependence arises in which the total island
density $N(t)\propto (\ln t)^{\mu/2}$.  In this regime, the island
distribution exhibits a rich mass dependence in which there is: (i) a
precipitous decay in a ``boundary layer'' $k\ll \kappa$ (with $\kappa\sim\ln
t$), (ii) a gradual growth in the main part of the mass distribution
$\kappa<k<K$ (with $K\sim t(\ln t)^{-\mu/2}$), and (iii) an internal layer
$|k-K|\sim \sqrt{t}$ where the density of islands reaches a peak and then
sharply vanishes.  The entire regime $1\leq \mu<\infty$ exhibits this same
behavior up to logarithmic corrections, while the transition between the
steady and evolving regimes at $\mu=1$ is characterized by nested logarithmic
behavior.

It is noteworthy that our theoretical approach can be applied to epitaxial
systems with {\em arbitrary} mass-dependent diffusivity $D_k$ which decays
faster than inverse mass.  For this general situation, the analog of
Eq.~(\ref{soln}) is
\begin{equation}
\label{ckdj}
c_k\sim N^{-1}\prod_{j=2}^{k}{1+D_j\over 1+N^2D_j}.
\end{equation}
For example, for a diffusivity which decays exponentially in island mass,
$D_k=e^{-a(k-1)}$, the case investigated numerically in Ref.~\cite{kuip}, our
theory predicts
\begin{equation}
\label{Nexp}
N(t)\sim \sqrt{F}
\exp\left[\sqrt{{a\over 2}\,\ln\left(t\sqrt{F}\right)}\right]  
\end{equation}
and $K\sim t/N$, leading again to the same universal value of the mass cutoff
exponent $\zeta(a)\equiv 1$.  Eq.~(\ref{Nexp}) exhibits unusual time
dependence -- faster than any power of logarithm and slower than any power
law -- and may be difficult to observe numerically.  The maximum island
density is
\begin{equation}
\label{Nexpmax}
N_{\rm max}\sim \sqrt{F}
\exp\left[\sqrt{{a\over 4}\,\ln(1/F)}\right],  
\end{equation}
so again $\chi=1/2$.  Numerically, the exponent $\chi(a)$ appears to decrease
as $a$ increases\cite{kuip}.  Our analysis suggests that in the asymptotic
regime $\chi(a)\equiv 1/2$ for all $0<a<\infty$.  However, fitting the
functional form in Eq.~(\ref{Nexpmax}) to a single power-law in $F$ gives
$\chi_{\rm eff}={1\over 2}-\sqrt{a/[4\ln(1/F)]}$.  Therefore even for small
flux the effective exponent may be considerably smaller than 1/2.  Also,
$\chi_{\rm eff}(a)$ is a decreasing function of $a$, in agreement with the
observations from the simulation\cite{kuip}.

\medskip\noindent
JFFM gratefully acknowledges support from FLAD and JNICT/PRAXIS XXI:
grant /BPD/6084/95 and project PRAXIS/2/2.1/Fis/299/94.  PLK and SR
gratefully acknowledge support from NSF grant DMR9632059 and ARO grant
DAAH04-96-1-0114.

\appendix
\section{Epitaxial Growth on 1d Substrates}

We now extend our results to the case of a one-dimensional substrate.  Since
the upper critical dimension $d_c=2$\cite{van}, the mean-field approximation
does not apply in one dimension.  In the absence of a theoretical framework
to systematically treat the case $d<d_c$, we give a heuristic treatment.  We
will derive results for the specific cases of $d=1$ and $d=2$; comparison
between the latter and the rate equation results provide a check of our
approach.

Consider first the simpler case of systems that approach a steady state.  We
present an argument based on the volume swept out by a Wiener sausage, as in
the case of systems without flux (Sec.\ IV).  To mimic the effect of the
flux, we suppose that there is no flux but that all islands have initial mass
which equals $t$.  At time $t$, all islands within a reaction volume
$(Dt)^{d/2}$ have coalesced into a single island.  Ignoring logarithmic
correction in two dimensions, this gives the following estimate for the
average island mass $M$,
\begin{equation}
\label{M}
M(t)\sim t\times\cases{Dt,        & $d=2$,\cr
                       \sqrt{Dt}, & $d=1$.\cr}
\end{equation}
Combining (\ref{M}) with $D\sim M^{-\mu}$ and using the fact that the average
island mass $M$ scales as the mass cutoff we find
\begin{equation}
\label{kc}
K(t)\sim \cases{t^{2/(1+\mu)}, & $d=2$,\cr
                  t^{3/(2+\mu)}, & $d=1$.\cr}
\end{equation}
In the steady state regime $c_k\sim k^{-\tau}$.  Consequently the sum rule
$t=\sum^K kc_k\sim K^{2-\tau}\sim t^{\zeta(2-\tau)}$ implies the
relation $\tau=2-1/\zeta$.  This, together with (\ref{kc}), gives
\begin{equation}
\label{tau}
\tau=\cases{(3-\mu)/2, & $d=2$,\cr
            (4-\mu)/3, & $d=1$.\cr}
\end{equation}
This argument reproduces the correct values of the cutoff and the decay
exponents when $d=2$, and we anticipate that the one-dimensional results are
also exact.  Indeed, an exact solution\cite{thom} of one-dimensional
aggregation with monomer input and mass-independent island diffusivities
gives $\tau=4/3$.  Finally, note that the exponents $\zeta$ and $\tau$ attain
the critical value $\zeta=\tau=1$ when the mobility exponent $\mu=1$ in both
one and two dimensions.

When $\mu\geq 1$, continuous evolution occurs.  We now argue that the
densities asymptotically evolve according to generalized rate equations.  For
monomers, we write
\begin{equation}
\label{mon}
{dc_1\over dt}=F-{c_1\over \Delta t}.
\end{equation}
Here $\Delta t$ is the collision time for a monomer to encounter an island.
During this time interval, a monomer visits $\sqrt{\Delta t}$ different sites
in one dimension, so that the collision time is determined by $N\sqrt{\Delta
  t}\approx 1$.  Consequently, the rate equation for the monomer density
becomes
\begin{equation}
\label{mono}
{dc_1\over dt}=F-N^2c_1.
\end{equation}
Continuing this line of reasoning we obtain the rate equations for the
densities $c_k(t)$ which differ from Eqs.~(\ref{smol}) by a factor $N$ in
each reaction term.  

We now analyze these equations by the same quasi-static framework as in Sec.\ 
II.  Thus we need to solve
\begin{eqnarray}
\label{1smol}
0&=&F-N^2c_1, \nonumber \\
0&=&\frac{1}{2}\sum_{i+j=k}\left({i^{-\mu}}+{j^{-\mu}}\right)c_ic_j
   -c_k\left(k^{-\mu}N+N_\mu\right).
\end{eqnarray}
Repeating the steps of our previous derivations we obtain, {\it e.g.}, for the
density of islands,
\begin{equation}
\label{1Nasymp}
N(t)\simeq F^{1/3}\left[{\sin(\pi/\mu)\over \pi}\,
\ln\left(tF^{2/3}\right)\right]^{\mu/2},
\end{equation}
and for the behavior of the density of relatively small islands
\begin{equation}
\label{1ck}
c_k(t)\sim F^{1/3}\,{(k!)^\mu\over N^{\mu k}}.
\end{equation}
This in the continuously evolving regime, the time dependence remains
primarily unaffected by the dimensionality of the substrate.  However, the
flux dependence does change with $d$, and we find $N_{\rm max}\sim
F^{1/3}[\ln(1/F)]^{\mu/2}$ (compared to the $F^{1/2}$ dependence in
Eq.~(\ref{Nmax})).  Overall, universal behavior arises in continuous
evolution which is only slightly affected by model details, substrate
dimensionality, {\it etc.}.

\section{Reaction Rate in Two Dimensions}

We now account for the logarithmic corrections to the reaction rate
(Eq.~(\ref{rate})) that appear in two dimensions.  We first demonstrate that
these logarithmic corrections can be accounted for within modified rate
equations.  Let us first consider a simpler model where point islands 
diffuse at the same mass-independent rate.  Then the total island density
$c(t)$ obeys $\dot c=-c/\Delta t$, where $\Delta t$ is the time between
successive collisions.  A collision is expected when the island visits $1/c$
distinct sites.  Since the number of distinct sites visited by a random walk
in time $t$ grows as $Dt/\ln(Dt)$ in two dimensions\cite{feller}, the
collision time follows from the condition $D\Delta t/\ln(D\Delta t)\sim 1/c$.
The resulting expression for $\Delta t$ leads to the rate equation $\dot
c=-Dc^2/\ln(1/c)$.  Similarly, for an island of radius $R$ we obtain $\dot
c=-Dc^2/\ln(1/cR^2)$.

For growing droplets with mass dependent diffusivity, these logarithmic
factors imply that the reaction rate $K_{ij}=D_i+D_j$ should be replaced by
\begin{equation}
\label{kij}
K_{ij}\sim {D_i+D_j\over \ln\left[N^{-1}(R_i+R_j)^{-2}\right]}.  
\end{equation}
In the low coverage limit, the average separation between neighboring islands
$N^{-1/2}$ is much larger than the average island size.  Keeping only this
dominant factor inside the logarithm gives the asymptotic form of the
reaction rate $K_{ij}\sim (D_i+D_j)/\ln(1/N)$.  Moreover, we can replace the
total island density by $\sqrt{F}$ inside the logarithm.  This is obvious
when $\mu<1$, since in this case the island density indeed approaches a
steady state value $N_\infty\sim \sqrt{F}$.  For $\mu\geq 1$, the island
density grows according to $N\sim \sqrt{F}(\ln t)^{\mu/2}$.  However, the
time-dependent factor is clearly sub-dominant as it is at most logarithmic in
the flux, $(\ln t_{\rm max})^{\mu/2}=(\ln(1/F))^{\mu/2}$.  Hence for
arbitrary mobility exponent $\mu$, the form $K_{ij}\sim (D_i+D_j)/\ln(1/F)$
provide an asymptotically correct description for the reaction rate.

Therefore in two dimensions we can continue to use the mean-field
Smoluchowski equations, with the modification of the reactive term by the
factor $1/\ln(1/F)$.  Upon rescaling the densities by $c_k\to
c_k\sqrt{F\ln(1/F)}$, and the time variable by $t\to t\sqrt{F^{-1}\ln(1/F)}$,
we formally map Smoluchowski equations for epitaxial growth in two dimensions
onto the mean-field Eqs.~(\ref{smol}) with $F=1$ and the reaction rates given by
(\ref{rate}).

We therefore conclude that we can apply the mean-field results to
two-dimensional substrates upon making the flux renormalization
\begin{equation}
\label{renorm}
F\to F\ln(1/F).
\end{equation}
This renormalization does not alter the basic predictions of the Smoluchowski
approach; for example, all exponents remain the same.  However, this
renormalization does alter some logarithmic factors, {\it e.g.},
Eq.~(\ref{Nmax}) for the maximum island density is replaced by
\begin{equation}
\label{Nmaxnew}
N_{\rm max}\sim F^{1/2}[\ln(1/F)]^{(\mu+1)/2}.
\end{equation}

\end{multicols}
\end{document}